# Performance degradation of Geiger-mode APDs at cryogenic temperatures

A. Bondar,[a,b] A. Buzulutskov,[a,b,*] A. Dolgov,[b] L. Shekhtman,[a,b] E. Shemyakina,[a,b] A. Sokolov,[a,b] A. Breskin[c] and D. Thers[d]

[a] *Budker Institute of Nuclear Physics SB RAS, Lavrentiev avenue 11, 630090 Novosibirsk, Russia*
[b] *Novosibirsk State University, Pirogov street 2, 630090 Novosibirsk, Russia*
[c] *Weizmann Institute of Science, 76100 Rehovot, Israel*
[d] *Laboratoire SUBATECH, UMR 6457 Ecole des Mines, CNRS/IN2P3 and Universite de Nantes, 4 rue Alfred Kastler, 44307 Nantes Cedex 3, France*

E-mail: `A.F.Buzulutskov@inp.nsk.su`

ABSTRACT: Two-phase Cryogenic Avalanche Detectors (CRADs) with THGEM multipliers, optically read out with Geiger-mode APDs (GAPDs), were proposed as potential technique for charge recording in rare-event experiments. In this work we report on the degradation of the GAPD performance at cryogenic temperatures revealed in the course of the study of two-phase CRAD in Ar, with combined THGEM/GAPD-matrix multiplier; the GAPDs recorded secondary scintillation photons from the THGEM holes in the Near Infrared. The degradation effect, namely the loss of the GAPD pulse amplitude, depended on the incident X-ray photon flux. The critical counting rate of photoelectrons produced at the 4.4 mm$^2$ GAPD, degrading its performance at 87 K, was estimated as $\sim 10^4$ s$^{-1}$. This effect was shown to result from the considerable increase of the pixel quenching resistor of this CPTA-made GAPD type. Though not affecting low-rate rare-event experiments, the observed effect may impose some limitations on the performance of CRADs with GAPD-based optical readout at higher-rate applications.

KEYWORDS: Cryogenic avalanche detectors (CRADs); Geiger-mode APDs at cryogenic temperatures.

---

[*] Corresponding author.

# Contents

## 1. Introduction

The performance of Geiger-mode APDs (GAPDs) (also known as SiPMs, MPPCs and MRS APDs [1]) at cryogenic temperatures is relevant in the field of Cryogenic Avalanche Detectors (CRADs), as reviewed in [2],[3]. In two-phase CRADs the primary ionization charge, composed of electrons produced in the noble liquid and emitted into the gas phase, can be multiplied in the gas phase with GEM [4] or THGEM [5],[6],[7] multipliers. In CRADs operated with combined THGEM/GAPD-matrix multipliers, the avalanche signal in the THGEM multiplier is optically recorded with a GAPD matrix [8],[9],[10]. In this mode, the GAPDs record secondary scintillation light from THGEM holes either in the VUV [10],[11] (using wavelength shifters) or in the Near Infrared (NIR) [8],[9],[12],[13],[14] spectral range. Such a combined charge/optical readout provides a superior spatial resolution [9],[10] at lower detection thresholds [11],[12] (see also table 2 in [2]). The development of CRADs with THGEM/GAPD readout has been motivated by their potential applications in low-threshold rare-event experiments; examples are direct dark matter searches, coherent neutrino-nucleus scattering and astrophysics neutrino detection [2],[3]. In particular, in a project presented recently by the Budker INP and Novosibirsk State University for dark matter search and low-energy neutrino detection [8],[15], a two-phase CRAD with THGEM/GAPD-matrix multiplier was proposed, with 1200 GAPDs sensors sensitive in the NIR and operated at 87 K.

It should be noted that though the GAPD operation at cryogenic temperatures was repeatedly studied [16],[17],[18],[19],[20],[21],[22],[23], the understanding of their performance at low temperatures is still incomplete. In particular, in this work we focus on a new effect, revealed in the course of the study of the two-phase CRAD in Ar with combined THGEM/GAPD-matrix multiplier [9], that of the pulse-amplitude degradation of the GAPDs operated at cryogenic temperatures. This effect may impose a significant limitation on the efficiency and rate capability of CRAD devices with GAPD-based optical readout.

## 2. Experimental setup

The experimental setup (see Fig. 1) was similar to those used in our previous studies of two-phase CRADs in Ar [7], [9]. It includes a cryostat with a 9 l volume cryogenic chamber. The chamber consisted of a cathode mesh, immersed in a ~1 cm thick liquid-Ar layer, and a double-THGEM assembly with an active area of $10\times10$ cm$^2$, placed in the gas phase above the liquid. The detector was operated in two-phase mode in the equilibrium state, at a saturated vapour pressure of 1.0 atm and at a temperature of 87 K. In this study, the charge gain of the double-THGEM multiplier was kept at rather moderate level, in the range of 40-160.

A 3x3 matrix of GAPDs (MRS APD "CPTA 149-35" [24], production of 2012) was placed in the gas phase at a distance of 6.5 mm behind the second THGEM, electrically insulated from the latter by an acrylic plate, transparent in the NIR, and by a wire grid at ground potential. Each GAPD had a $2.1\times2.1$ mm$^2$ active area with 1764 pixels, a capacitance of 150 pF and a Photon Detection Efficiency (PDE) of about 15% at 800 nm [12],[24]. The gain-voltage and noise-rate characteristics of this particular GAPD type, at 87 K, were presented elsewhere [22]. In practically all the measurements at 87 K the GAPD bias voltage was 40 V, providing a typical gain of about $1\times10^6$ and noise rate of about 20 s$^{-1}$.

The detector was irradiated from outside through aluminium windows by 15-40 keV X-rays from a pulsed X-ray tube with Mo anode operated at 40 kV, with a pulse rate of 240 s$^{-1}$, through a cylindrical collimator with a hole diameter of 2 mm, similarly to that of [9]. The X-ray conversion in the liquid was defined by this collimator and occurred over a small area of ~2 mm in diameter: see Fig. 2 showing the appropriate reconstructed X-Y plot. Here the squares with numbers indicate the positioning of the active areas of the appropriate GAPDs in X-Y plane. The coordinate reconstruction algorithm is described elsewhere [9].

There were two sets of measurements performed at different distances to the X-ray tube and with the presence of X-ray attenuation filters. In the first run, without intermediate Al filters, the X-ray tube was placed at a relatively large distance, of ~50 cm from the collimator (Fig. 1), providing either one or a few X-ray photons per pulse with an average deposited energy of about 20 keV per pulse. Accordingly, in these conditions the X-ray flux was close to the tube's pulse rate, i.e. $\geq240$ s$^{-1}$.

In the second run, a set of intermediate Al filters (plates) were placed between the X-ray tube and the collimator, to attenuate the incident X-ray flux and thus to vary the counting rate. This permitted studying the incident-flux dependence of the pulse-degradation effect. In this run, the X-ray tube was placed substantially closer to the collimator. The absolute value of the X-ray flux, at a minimal flux, prior to an observable degradation effect, was determined by measuring the counting rate of the two-phase CRAD. At higher fluxes the absolute flux values were calculated using a dedicated computer program for a given X-ray tube anode, anode voltage and Al filter thickness. Accordingly, the incident X-ray flux varied from 10 to $1.2\times10^4$ s$^{-1}$, the X-ray photon energy being in the range of 15-40 keV with the average deposited energy of about 20 keV.

The charge signals were recorded from the last electrode of the second THGEM using a charge-sensitive amplifier. The optical signals were recorded by the GAPDs and transferred via twisted-pair cables to fast amplifiers with 300 MHz bandwidth and an amplification factor of 30. The DAQ system included an 8-channel Flash ADC CAEN V1720 (12 bits, 250 MHz): the signals from 7 GAPDs and from the THGEM were digitized and stored in a computer for further off-line analysis using a LabView program [25] (i.e. only 7 out of the 9 GAPDs were actively recorded).

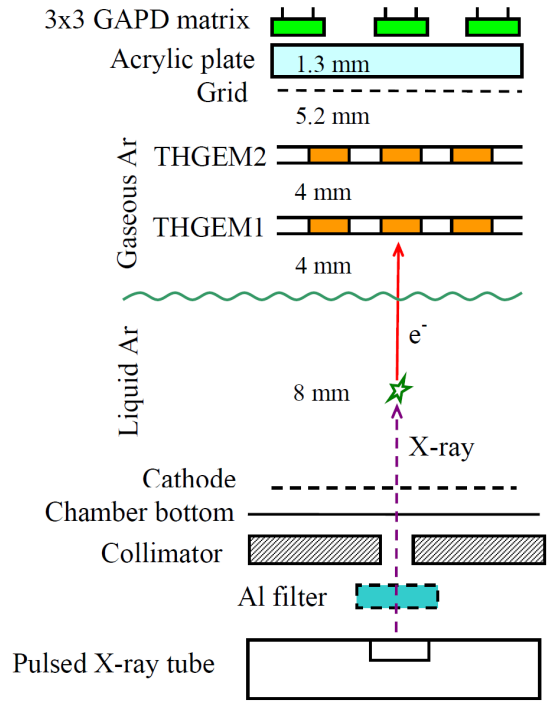

Fig. 1. Experimental setup.

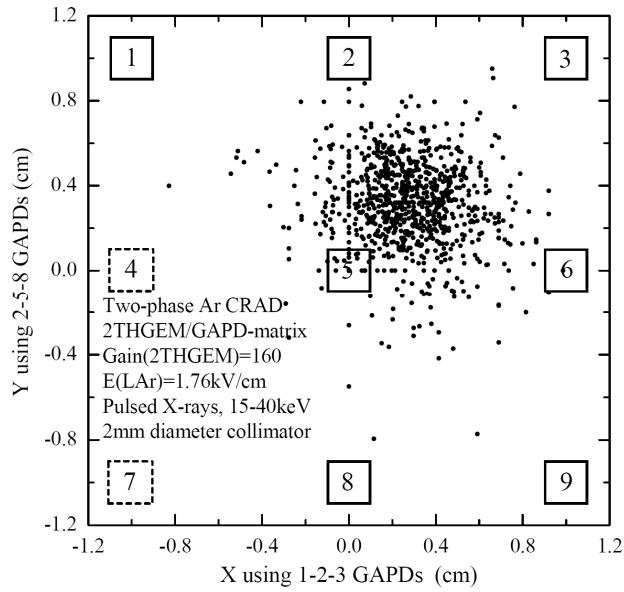

Fig. 2. X-Y plots of the pulsed X-ray conversion region in the two-phase CRAD in Ar with a THGEM/GAPD-matrix multiplier, defined by a 2 mm diameter collimator (1st run). Squares with numbers indicate the positioning of the active areas of the appropriate GAPDs.

The trigger was provided by the pulsed X-ray tube generator. In this case the contribution of noise signals was negligible. Other details of the experimental setup and procedures are provided elsewhere [7],[9],[12],[13].

## 3. GAPD performance degradation at 87 K

At 87 K, with no X-ray irradiation and with a GAPD noise level of a few tens of Hz, its pulse amplitude distribution had a nominal shape: a well-defined single-pixel peak at about 15mV, well separated from zero, accompanied by secondary (cross-talk) peaks. Fig. 3, shows the noise pulses and their pulse-height distributions for GAPDs #5 (central) and #8 (peripheral), located respectively near and far from the X-ray conversion region (see Fig. 2).

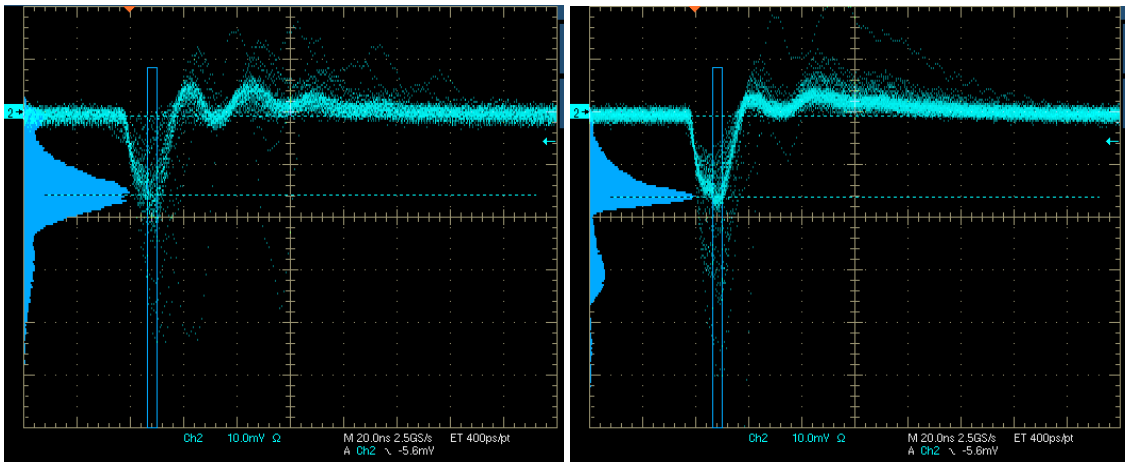

Fig. 3. Noise pulses and their pulse-height distributions of GAPD #5 (left) and #8 (right), recorded in the two-phase CRAD in Ar with a THGEM/GAPD-matrix multiplier. Scales: 20 ns/div; 10 mV/div. GAPD bias voltage: 40 V.

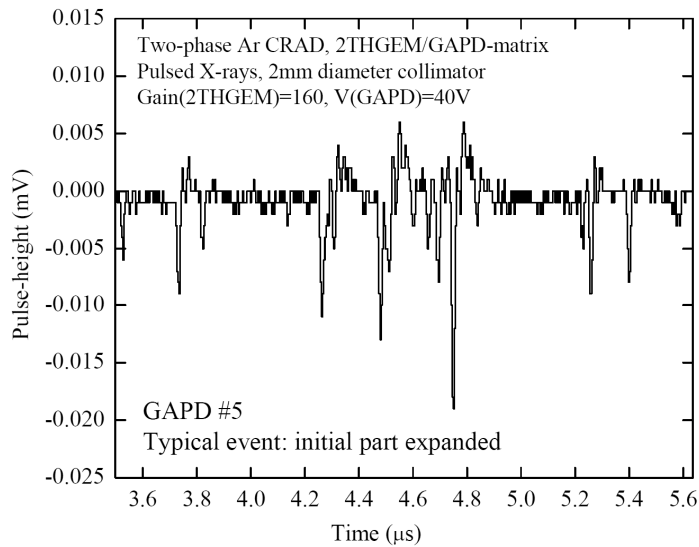

Fig. 4. Initial part of a typical signal of GAPD #5 recorded in the two-phase CRAD in Ar with a THGEM/GAPD-matrix multiplier; they were induced by pulsed X-rays at a rate of 240 s$^{-1}$ with an average deposited energy of 20 keV per pulse (1st run, without Al filters). Double-THGEM gain: 160; GAPDs bias voltage: 40 V.

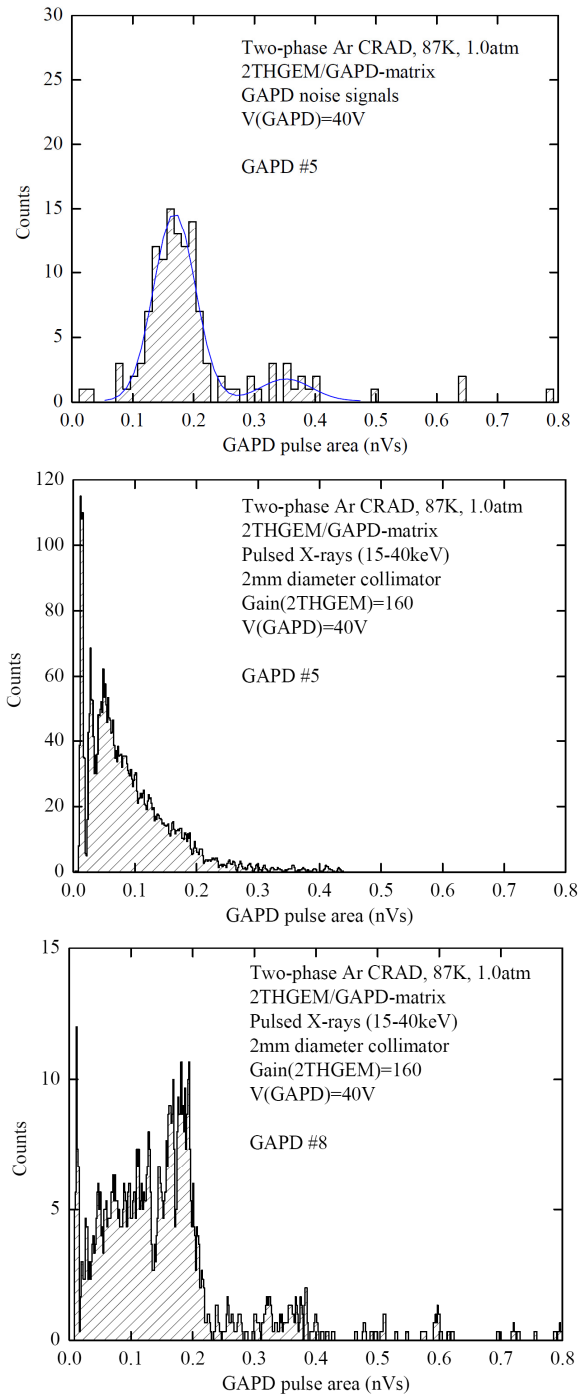

Fig. 5. Pulse-area distributions of GAPD #5 (middle) and #8 (bottom) signals in the two-phase CRAD in Ar with a THGEM/GAPD-matrix multiplier, induced by pulsed X-rays at a rate of 240 s$^{-1}$ with an average energy of 20 keV per pulse (1st run, without Al filters). For comparison, that of the GAPD #5 noise signals (with no X-ray irradiation), is shown (top). Double-THGEM gain: 160; GAPDs bias voltage: 40 V.

However, at higher rates, when irradiated with the pulsed X-ray tube during the 1st measurement run, a specific performance degradation effect has been observed: the pulse-height of the GAPDs substantially decreased, sometimes down to 5 mV and below. This is seen in Fig. 4 showing the initial part of a typical signal of GAPD #5 in the two-phase CRAD, induced by pulsed X-rays at a rate of 240 $s^{-1}$ with an average deposited energy of 20 keV per pulse. Here the double-THGEM charge gain was 160. One can see that the GAPD signal is composed of a number of fast pulses, each apparently corresponding to a single-pixel discharge (or those of multi-pixels in case of cross-talk), i.e. to the detection of a single photoelectron per pixel. It should be remarked that the rather slow signals observed here in the two-phase CRAD, sometimes exceeding 20 μs, are due to the slow electron emission component present in two-phase Ar systems [26]. Note that some pulses could be even lost below detection threshold, thus apparently reducing the GAPD efficiency.

The performance degradation was also noticeable in the amplitude spectra; Fig. 5 shows the pulse-area distributions of GAPD #5 and #8 signals, induced by pulsed X-rays at a rate of 240 $s^{-1}$ (1st run). The spectrum degradation is distinctly seen in particular when comparing these spectra to that of GAPD #5 noise signals (with no X-ray irradiation): while the latter has a nominal shape with well-defined single- and double-pixel peaks, the amplitudes of the former tends to shift to zero. It is interesting that this degradation effect is more and less pronounced for GAPD #5 and #8, respectively; apparently it is due to the correspondingly higher and lower incident flux of the THGEM-induced NIR photons. This confirms the dependency of the degradation effect on the incident NIR photon flux at a particular photodiode.

This flux dependency was studied quantitatively in the 2nd run, using Al filters to vary the X-ray flux: see Fig. 6. One can see that the average pulse-area amplitude of the GAPD #5 signals substantially decreases (by 25%) at X-ray photon flux as low as 250 $s^{-1}$, with an average X-ray photon energy of 20 keV. The more monotonic decrease at higher fluxes is explained by the fact that the major part of the amplitude spectrum is now below the detection threshold; as a result, the average amplitude of the remaining events becomes less sensitive to the photon flux.

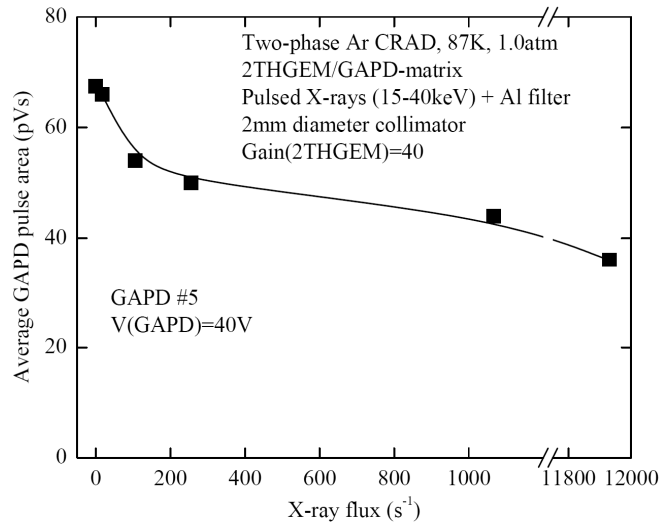

Fig. 6. Average pulse-area of GAPD #5 signals as a function of the incident X-ray photon flux in the two-phase CRAD in Ar with THGEM/GAPD-matrix multiplier (2nd run, with Al filters). Average X-ray photon energy: 20 keV; double-THGEM gain: 40; GAPDs bias voltage: 40 V.

From this value we can roughly estimate the critical rate of photoelectrons produced in the GAPD at which its performance starts degrading. We took into account the following factors: the number of ionization electrons prior to multiplication in the THGEM multiplier (~1000 electrons per 60 keV X-ray [6] and correspondingly ~300 electrons per 20 keV); the double-THGEM gain (40), the NIR photon yield per avalanche electron (4 photons/$e^-$/$4\pi$, see table 2 in ref. [2]); the reduced solid angle of the GAPD with respect to the light emission region in the holes of the second THGEM ($\Delta\Omega/4\pi$=0.0045 using Monte-Carlo simulation); and the GAPD Photon Detection Efficiency (PDE) in the NIR in the Ar emission range (15% at 800 nm). These lead to the value of ~40 photoelectrons produced in GAPD #5 per incident X-ray photon. That bring us to the conclusion that the critical counting rate of photoelectrons produced at the GAPD, degrading the performance observed at 87 K, is of the order of $40 \times 250 = 1 \times 10^4$ $s^{-1}$.

## 4. GAPD quenching resistor

We suppose that the GAPD performance degradation at cryogenic temperatures results from the increase of the GAPD pixel recovery time $\tau = R_Q C_P$, due to the increase of the pixel quenching resistor $R_Q$ at cryogenic temperatures; here $C_P$ is the pixel capacitance. It should be remarked that the increase of $R_Q$ at low temperatures was observed earlier [12].

In order to estimate $R_Q$, the GAPD current-voltage characteristics were measured in the forward direction at different temperatures; Fig. 7 shows these characteristics for two different GAPDs of different production batches – of 2009 and 2012. The slope of the I-V curve in its linear part corresponds to the total quenching resistor value R, being the sum of 1764 pixel quenching resistors connected in parallel; thus the pixel quenching resistor value is $R_Q=1764 \times R$. In addition, Fig. 8 illustrates the distribution of the total quenching resistor value at room temperature over the batch of 9 GAPDs from the 2012 production, i.e. over those studied in the present work. Finally, the dependence of the quenching resistor value on the temperature is demonstrated in Fig. 9: the value $1/R_Q$ is shown as a function of $1/T$; in these variables, the Boltzmann-type temperature dependence of the conductivity $\sigma \sim \exp(-E_a/k_B T)$ (with $E_a$ being the activation energy of the conductivity and $k_B$ the Boltzmann constant) is better revealed. Analyzing the data in Figs. 7-9, one may conclude the following.

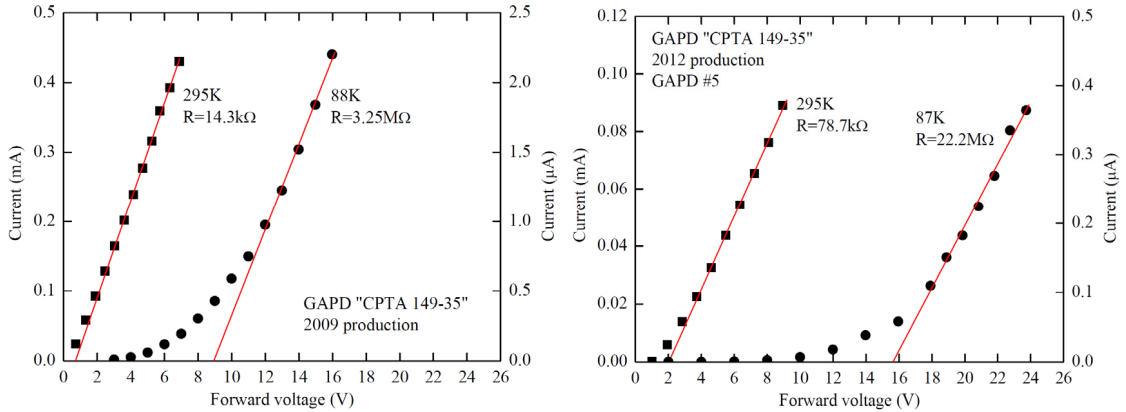

Fig. 7. GAPD current-voltage characteristic in the forward direction at room temperature (left scale) and at 87 K (right scale). The slope of the linear part of the I-V curve is defined by the GAPD total quenching resistor R. Left: GAPD of the 2009 production. Right: GAPD #5 of the 2012 production.

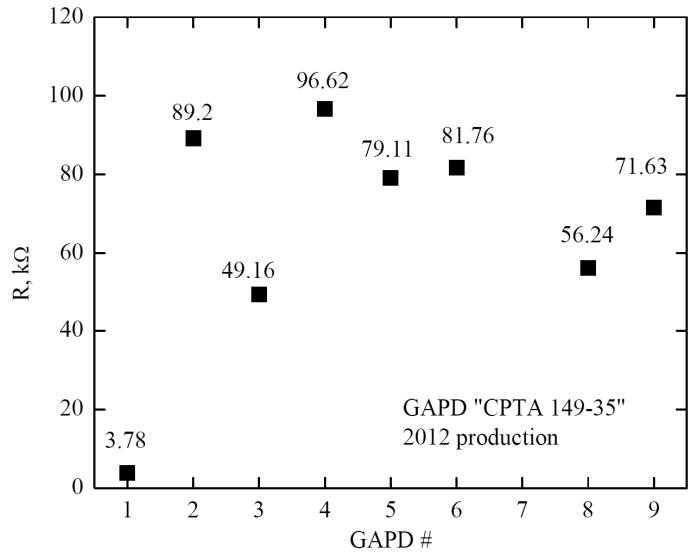

Fig. 8. Total quenching-resistor values of the 9 CPTA-GAPDs of the 2012 production, measured at room temperature.

The first conclusion is that the quenching resistor dramatically increases with temperature decrease, by more than two orders of magnitude: in particular for GAPD #5 the pixel quenching resistor changes from 140 MΩ at room temperature to 40 GΩ at 87 K (see Figs. 7 and 9). From this we can estimate the rate capability of the GAPD at 87 K. Taking into account the pixel capacitance $C_P$=150pF/1764=85 fF, the pixel recovery time is estimated to be of the order of $\tau$=10 µs and $\tau$=3 ms at room temperature and at 87 K respectively. Since the pixel voltage recovers as V~1-exp(-t/$\tau$), one should have as much as say 10$\tau$ for almost full recovery. Accordingly, the pixel rate capability at 87 K is somewhat lower than the value 1/10$\tau$=30 s$^{-1}$. For the entire GAPD this limit is increased somewhat slower than in proportion to the number of pixels (due to the probability to hit the same pixel), i.e. it is lower than 1764/10$\tau$=5×10$^4$ s$^{-1}$. This does not contradict with the experimental limit presented in the previous section, of 1×10$^4$ s$^{-1}$.

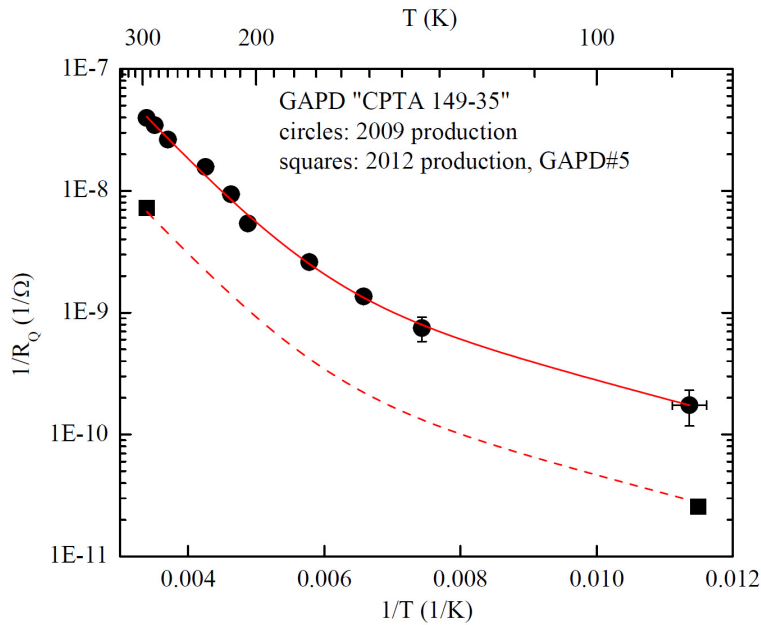

Fig. 9. GAPD single-pixel quenching-resistor value ($R_Q$) dependence on temperature: the reciprocal value $1/R_Q$ is shown as a function of the reciprocal temperature $1/T$ for two CPTA-GAPDs of the 2009 and 2012 production batches.

The second conclusion is that the GAPD quenching resistors have a large scatter within the batch, at least by a factor of 2 in the 2012 production (see Fig. 8). Moreover, compared to the resistors of the 2009 production batch, those of 2012 have considerably higher values, by a factor of 4-7. Such variations in the production characteristics of the CPTA-GAPDs are rather disturbing.

The third conclusion is that the higher the quenching resistor value at room temperature, the higher is that at cryogenic conditions; i.e. the temperature dependence of the quenching resistors of different GAPDs have similar behavior: see Fig. 9.

In addition, we did not observe the degradation effect at cryogenic temperatures for the GAPD of the 2009 production, having a factor of 4-7 lower quenching resistor value compared to that of the 2012 one.

Finally, these findings support the hypothesis of the decisive effect of the quenching resistor on the GAPD rate capability at cryogenic temperatures.

5. **Summary**

In this work a new effect in GAPD performance at cryogenic temperatures is described, that of GAPD pulse-amplitude degradation when operated in a two-phase cryogenic avalanche detector (CRAD) in Ar at 87 K. The GAPDs were recording NIR secondary scintillation photons from THGEM-hole avalanches. The degradation effect turned out to be dependent on the incident photon flux, when the CRAD was irradiated with pulsed X-rays. The critical counting rate of photoelectrons produced at the GAPD (having an active area of 4.4 mm$^2$) to degrade its performance at 87 K, was estimated to be of the order of 1×10$^4$ s$^{-1}$.

This degradation effect was shown to result from the considerable increase of the pixel quenching resistor, at cryogenic temperatures, of the GAPDs manufactured by CPTA. The quenching-resistor values of the CPTA-GAPDs varied considerably within and between the production batches. The degradation effect is being investigated by us on other GAPD types.

The observed degradation may impose limitations on the performance of CRADs with GAPD-based optical readout at high-rate applications, e.g. in medical imaging systems. On the other hand, this degradation effect will presumably not be a problem for rare-event experiments, such as dark-matter searches - running at very low overall rates, even during their calibration runs with external radioactive sources. E.g. in XENON100, as well as in the forthcoming XENON1t experiment, the maximum rate in calibration runs with gamma sources is ~0.15 s$^{-1}$/cm$^2$ [27],[28]. This would roughly correspond to 100 s$^{-1}$ of photoelectrons produced at the GAPD in conditions of the present work (at a double-THGEM gain of 100), which is below the critical rate.

## 6. Acknowledgements

This work was supported in part by the Ministry of Education and Science of Russian Federation and by the grants of the Government of Russian Federation (11.G34.31.0047) and the Russian Foundation for Basic Research (12-02-91509-CERN_a).